\documentstyle[12pt]{article}
\def \thesection {\arabic{section}.}
\def \sect #1 {\setcounter{equation} 0\section{#1}}
\def \theequation {\thesection\arabic{equation}}
\def \appendix #1#2 {\par\noindent
                    {\Large {\bf Appendix {#1}. {#2} }}
                    \def\theequation{{#1}.\arabic{equation}}
                    \setcounter{equation} 0 \par\bigskip\noindent}
\def \figures {\begin{center}{\Large {\bf Figures: }}\end{center}\par\bigskip}
\def \be  {\begin{equation}}
\def \ee  {\end{equation}}
\def \bb  {\begin{thebibliography}}
\def \eb  {\end{thebibliography}}
\def \lab #1 {\label{#1}}
\newcommand \ci [1] {\cite{#1}}
\newcommand \bi [1] {\bibitem{#1}}
\newcommand\re[1]{(\ref{#1})}

\begin{document}
\def\thefootnote{\fnsymbol{footnote}}
\thispagestyle{empty}
\hfill\parbox{35mm}{{\sc ITP--SB--95--19}\par
                         May, 1995}
\vspace*{40mm}
\begin{center}
{\LARGE The soft pomeron and nonperturbative corrections \\[5mm]
in quark-quark scattering.}
\par\vspace*{20mm}\par
{\large Irina A. Korchemskaya}
\footnote{Irina@max.physics.sunysb.edu; on leave from the Moscow
          Energy Institute, Moscow, Russia}
\par\bigskip\par\medskip
{\em Institute for Theoretical Physics, \par
State University of New York at Stony Brook, \par
Stony Brook, New York 11794 -- 3840, U.S.A.}
\par\medskip
{\em and}
\par\medskip
{\em Universit\'a di Parma and INFN, \par
Gruppo Collegato di Parma, I--43100 Parma, Italy}

\end{center}
\vspace*{20mm}

\begin{abstract}
We consider elastic quark-quark scattering at high energy with fixed
momentum transfer $t$ and perform factorization of soft-gluon exchanges into
a vacuum expectation value of Wilson lines.
Taking into account nonperturbative corrections
whose structure is predicted from infrared renormalon analysis,
we represent the scattering amplitude as an asymptotic series. In the
region of small momentum transfer, where the nonperturbative corrections
are dominant, the scattering amplitude is Gaussian
distribution in $t$ with a slope depending on a
nonperturbative scale. A nonperturbative origin of the soft pomeron is thus
identified.
\end{abstract}
\newpage
\section{Introduction}
\setcounter{equation} 0
Regge theory \ci{Regge} explains a large class of experimental results
for hadronic scattering amplitudes at high energy $s$ and fixed transferred
momentum $t$.
After almost 35 years, however, it remains a challenge to understand Regge
theory within the framework of fundamental quantum field theory. One
such attempt led to the development of the BFKL pomeron \ci{BKFL}, which
describes the compound state of two reggeized gluons
with vacuum quantum numbers.
Found in the leading logarithmic approximation, the BKFL pomeron,
sometimes called the hard pomeron, leads to scattering
amplitudes which violate the Froissart bound. Recently
\ci{Lip}, \ci{FK}, \ci{K.BA for Pomeron}
it was shown that in generalized leading logarithmic approximation
QCD is described by an effective 2-dimensional field
theory, which is equivalent to an XXX Heisenberg magnet of spin $s=0$.
This theory takes into account the propagation of infinite
numbers of interacting Reggeons in the $t$-channel and restores the unitary of
the S-matrix.

This paper is devoted to the ``soft'' pomeron \ci{Landshoff. Two Pomerons}
, \ci{Bertini.Giffon.Predazzi}.
While this idea is phenomenologically
very successful \ci{Donnachie.Landshoff.old}, and we know from experiment
the pomeron trajectory, $\alpha(t)=1.08 + 0.25t$,
a deep understanding how the pomeron appears in QCD is still lacking.
It is widely believed that the soft pomeron has a nonperturbative origin.
It was proposed in \ci{nonpert.cor.} to include nonperturbative effects
through a modification of the gluon propagator. Another ideas, in particular
the application of the method of the stochastic vacuum models
were discussed in \ci{Nachtmann}. In \ci{Verlinde} the high energy
interactions were described in terms of a two-dimensional sigma-model action.
The approach developed here follows from a different perspective.
We shall exploit the viewpoint that perturbation
theory itself can predict the form of nonperturbative corrections,
and show that ambiguities of the
perturbative series caused by infrared renormalons allow one
to identify the structure of nonperturbative corrections.
In case of $e^{+}e^{-}$ annihilation, which admits the operator product
expansion, the nonperturbative corrections can be parameterized by
local vacuum condensates. Further analysis \ci{Mueller} revealed that
the perturbative series is not Borel summable. The singularities
of the Borel transform, which are called IR renormalons \ci{t'Hooft}
imply that the physical quantity
will be well defined if the ambiguity caused by IR renormalons is
compensated by an ambiguity in the definition of local vacuum condensates.
In \ci{kor.ster} the idea of IR renormalons was generalized
to hadronic processes (jet cross sections, inclusive Drell-Yan
lepton-pair production) to which the operator product expansion is not
applicable. It was found that nonperturbative corrections are
parameterized by new parameters, associated with vacuum expectation values
of nonlocal operators involving Wilson lines and the gluon field strength.
The strength of the nonperturbative corrections is determined by the position
of the leading renormalon. The relation between infrared renormalons and
power corrections was also discussed in \ci{CS}, \ci{Bigi}, \ci{Neubert},
\ci{Manohar.Wise}, \ci{Dokshitzer.Webber}.

Analyzing high-energy hadronic scattering, we consider hadrons as
consisting of partons. The soft pomeron is thought to couple to
the valence quarks only.
This additive quark rule is supported by much experimental data.
Quark-quark scattering naturally involves into consideration
the Wilson lines.
In nonperturbative QCD the Wilson line appears, for example, in the
representation of the quark propagator $S(x,y;A)$  as a sum over random paths
between points x and y \ci{Polyakov},\ci{Spinfactor} and it takes into
account the interactions of gauge field with the color current created by the
quark moving along the path. The remarkable fact, however, is that in
high energy scattering, $s \gg -t$, the quarks move along the straight lines.
In perturbative QCD the Wilson lines describes the infrared asymptotics of the
quark propagator \ci{Kor.Rad}. Moreover one can expand the Wilson line
in powers of gauge fields and reproduce the eikonal approximation for the
interaction vertices of quark with soft gluons.

Our strategy is the following.
We start with factorization of the soft gluon exchanges into a vacuum
expectation value of Wilson lines and
represent the quark-quark scattering amplitude
as an expectation value of a Fourier transformed Wilson
line, evaluated along an integration path which consists of two
semiclassical quark trajectories, separated by an impact parameter in
the transverse direction \ci{Korchemsky}. For the case in which the
quarks are described by light-like Wilson lines the similar formula was
proposed by Nachtmann \ci{Nach.old}.
We shall go on and find the expression for the quark-quark
elastic scattering amplitude, which resembles an eikonal formula of
Cheng and Wu \ci{Cheng.Wu}.
We shall derive it, however, from the
renormalization properties of the cross singularities of Wilson loops
\ci{Korchemsky}.

Calculated to the lowest order of perturbation theory the
quark-quark scattering amplitude gets large perturbative corrections
such as $(\alpha_{s}\log s\log t)^{n}$.
Resummation of these Sudakov corrections
can be performed using
an  ``evolution equation'' technique \ci{Lip.quaseilastic},
\ci{Sotirop.Sterman}.
In the present paper, we generalize the method \ci{Korchemsky} to
perform the resummation of both kind of corrections: perturbative and
nonperturbative in the impact parameter space.
The perturbative Sudakov corrections come from very soft virtual gluons
with transverse momenta much smaller than the hadronic scale.
In this region the nonperturbative corrections
are very large. As we will show below, they have the form of power corrections
$(\rho\Lambda_{QCD}^2b^2)^n$, where $\rho$
is a parameter characterizing the nonperturbative interactions, while the
perturbative corrections behave as $(\alpha_{s}\log b^2)^n$.
The dependence of the scattering amplitude on
$t$ comes from the dependence of the Wilson lines on the impact parameter $b$.
This means that we expect a more or less realistic prediction for the $t$
dependence of the scattering amplitude, and our soft pomeron predicts a
linear Regge trajectory. It is problematic to find the intercept
of the soft pomeron from our model and we can only say that it is
close to unity.

The paper is organized as follows. In Section 2 we review
the application of the Wilson loop formalism to quark-quark scattering
\ci{Korchemsky},\ci{Kor.Kor}. In Section 3 we show that the
Wilson loop in perturbation theory contains an ambiguous
contribution to power corrections caused by infrared renormalons. In Section 4
using the prediction for the asymptotics of the cross anomalous
dimension \ci{Kor.Kor} in all orders of perturbation theory
we include the nonperturbative corrections in the expression for the
scattering amplitude, and represent the scattering amplitude as a Mellin
integral. In Section 5 we calculate the scattering amplitude for the
case of a frozen coupling constant. In Section 6 we consider the case of a
running coupling constant and represent the scattering amplitude as an
asymptotic series. Here we discuss the origin of the asymptotic expansion
as a consequence of the presence of a singularity in the running coupling
constant. We also derive the main results of this analysis: a
Gaussian distribution over transferred momenta in the scattering amplitude,
shrinkage of the distribution with increasing energy, and a crossover region in
the differential cross-section.
Section 7 contains concluding remarks.

\section{Scattering amplitude in perturbative QCD}
\setcounter{equation} 0
We consider near forward elastic quark-quark scattering at high energy  and
fixed transferred momentum in the following kinematics:
$$
s,\ m^2 \gg -t \gg \lambda^2 \geq \Lambda^2_{QCD}\,.
$$
Here  $s=(p_{1}+p_2)^2$  is the invariant energy of quarks with mass $m$
, $ t=(p_1-p_1')^2$  is the transferred momentum and $\lambda^2$ is an
IR cutoff. Let us explain the origin of this kinematics.
The quark-quark scattering should be thought of as embedded
in a physical process involving hadron-hadron elastic or inelastic scattering.
Considered in isolation, the q-q scattering amplitude has IR divergences and
an IR cutoff is necessary. In exclusive and inclusive physical processes
involving the q-q scattering, the IR divergences are canceled and the
IR cutoff is replaced by a dynamically generated transverse momentum scale
of the hadronic state. Therefore for a purely perturbative calculation
we should choose $\lambda^2 \gg \Lambda^2_{QCD}$. To include nonperturbative
effects, however, we find it useful to relax this condition such as
$\lambda^2 \geq \Lambda^2_{QCD}$.
As a particular example of hadronic state one can consider the
perturbative onium state \ci{Mul.onium} built from heavy quarks
with mass $m$. In this case the mass $m$ has a meaning of the transverse
size of the hadron.

In the center of mass frame of the incoming quarks, quark momenta have the
following light-cone components:
$p_1=m/\sqrt{2}(e^{\gamma/2}, e^{-\gamma/2}, \vec 0)$ and
$ p_2=m/\sqrt{2}(e^{-\gamma/2}, e^{\gamma/2}, \vec 0)$. In the limit
$s \gg m^2$ the angle $\gamma$ between quark velocities become large
and both quark move close to the ``+'' and ``-'' light-cone directions. The
components of the total momentum transfer $k=p_{1}-p_{1}'$ are
$k^{+}=-k^{-}=O(t/ \sqrt s)$ and $\vec k^2=O(t)$. Thus, in the limit
$s \gg -t$ we can neglect the longitudinal components of transferred
momentum and put $k=(0^{+}, 0^{-}, \vec k)$.
Since the transferred momentum is much smaller than the energies of
incoming quarks, the quarks interact each other by exchanging soft gluons
in the $t$-channel with total momentum $k$.
Interacting with each of soft gluons quark does not alter its velocity
in the limit $-t \ll m^2$ and thus the only effect of its interaction is
the appearance of an additional phase in the quark wave function.
This phase, the so-called eikonal phase, is equal to a Wilson line
$P\exp(i\int_{C} dx_{\mu}A_{\mu}(x))$ evaluated along the classical
trajectory  $C$ of quark in the direction of the quark velocity.
We combine the eikonal phases of both quarks and obtain the representation
for the scattering amplitude as \ci{Korchemsky}:
\be
T_{ij}^{i'j'}(\frac{s}{m^2},\frac{k^2}{\lambda^2})=\sinh\gamma\int d^2b\,
\\e^{-i\vec b\cdot \vec k}W_{ij}^{i'j'}
(\gamma,\vec b^2\lambda^2)  \, , \qquad\quad  t=-{\vec k}^2,
\lab{main}
\ee
where the line function $W_{ij}^{i'j'}$ is given by
\be
W_{ij}^{i'j'}(\gamma,\vec b^2\lambda^2)=\langle 0| T
\left[P\exp\left(ig\int_{-\infty}^\infty d\alpha\, v_1\cdot A(v_1\alpha)
\right)\right]^{i'}_{i}
\left[P\exp\left(ig\int_{-\infty}^\infty d\beta\, v_2\cdot A(v_2\beta+b)
\right)\right]^{j'}_{j}
|0\rangle.
\lab{main1}
\ee
Here the line function $W_{ij}^{i'j'}$ contains color indices of both incoming
$(i,j)$ and outgoing $(i',j')$ quarks.
The two Wilson lines are defined in the fundamental representation of
the SU(N) gauge group and evaluated along infinite paths in the direction
of the quark velocities $v_1=p_1/m$ and $v_2=p_2/m$. The integration
paths are separated by impact vector $b=(0^{+},0^{-},\vec b)$ in the transverse
direction, $v_1\cdot b=v_2\cdot b=0$.

The scattering amplitude \re{main} depends on the quark velocities $v_1$ and
$v_2$, the transferred momentum $k$ and the IR cutoff $\lambda$. These
variables give rise to only two scalar dimensionless invariants: $(v_1v_2)=
s/m^2$ and $t/\lambda^2$, as explicitly indicated in \re{main}.
The $s$-dependence of the amplitude comes from the dependence on the
angle $\gamma$ between quark four-velocities $v_1$ and $v_2$, defined in
Minkowski space-time as
$
(v_1v_2)=\cosh\gamma,
$
while its $t$-dependence is related to the dependence of the line function
on the impact vector $b$.
In the limit of high-energy quark-quark scattering $ (s \gg m^2)$ we have
\be
\gamma=\log\frac{s}{m^2} \gg 1\,.
\lab{g1}
\ee
The line function $W_{ij}^{i'j'}$ is divergent for $b=0$. This divergence,
the so called cross divergence, has an ultraviolet origin, because for $b=0$
the integration paths of Wilson lines (See Fig. 1a) cross each other.
According to the general analysis in \ci{Brandt},
the Wilson line  $(W_1)_{ij}^{i'j'}$ of Fig. 1a is mixed
under renormalization with the Wilson line  $(W_2)_{ij}^{i'j'}$ of
 Fig. 1b. As a consequence, the renormalized line functions $W_1$  and $W_2$
satisfy the following renormalization group equation:
\be
\left(\mu\frac{\partial}{\partial \mu}+\beta(g)\frac{\partial}{\partial g}
\right)W_a=-\Gamma_{\rm cross}^{ab}(\gamma,g)W_b   \qquad\quad a,b=1,2,
\lab{RG}
\ee
where $W_a\equiv (W_a)_{ij}^{i'j'}$ and $\mu$ is the renormalization scale.
Here, $\Gamma_{\rm cross}^{ab}$ is the cross anomalous dimension, which is a
gauge-invariant $2\times2$ matrix, depending only on the coupling constant and
the angle $\gamma$ between the lines at the cross point. Solving the RG
equation
for $W_1(\gamma,\frac{\mu^2}{\lambda^2})$  with the boundary conditions
$W_1(\gamma,1)=\delta_{i'i}\delta_{j'j}$ and $W_2(\gamma,1)=
\delta_{j'i}\delta_{i'j}$ and identifying the UV cutoff $\mu^2$ with $1/b^2$ we
find the following expression for the scattering
amplitude \ci{Korchemsky}, \ci{Kor.Kor}:
\be
T_{ij}^{i'j'} \left(\frac{s}{m^2}, \frac{k^2}{\lambda^2}\right)
=\sinh\gamma\int d^2b\,\\e^{-i\vec b\cdot \vec k}(A_{11}(\gamma,b^2\lambda^2)
\delta_{i'i}\delta_{j'j}
+ A_{12}(\gamma,b^2\lambda^2)\delta_{j'i}\delta_{i'j})\,,
\lab{main2}
\ee
where $A_{11}$  and  $A_{12}$  are elements of the $2\times2$ matrix
\be
A(\gamma,b^2\lambda^2)= \rm T\exp\left(-\int_{\lambda}^{1/b}
\Gamma_{\rm cross}(\gamma,\alpha_s(\tau))\frac{d\tau}{\tau}\right)\,.
\lab{A}
\ee
We conclude that
the asymptotic behavior of the scattering amplitude is governed by the matrix
of the cross anomalous dimension $\Gamma_{\rm cross}(\gamma,\alpha_s)$.
The expression \re{A} takes into account not only all  $\log t$  and
$\log s$  corrections but nonperturbative corrections as well.
As follows from \re{A}, the $t$-dependence
originates from the evolution of the coupling constant.

The one-loop expression for the matrix cross anomalous dimension is
$$
\Gamma_{\rm cross}(\gamma,g)=\frac{\alpha_s}{\pi}\Gamma_{\rm cross}(\gamma),
$$
\be
\Gamma_{\rm cross}(\gamma)=\left(\matrix{
-\frac{i\pi}{N}\coth\gamma        &i\pi\coth\gamma\cr
-\gamma\coth\gamma+1+i\pi\coth\gamma
&N(\gamma\coth\gamma-1)-\frac{i\pi}{N}\coth\gamma
\cr}\right)\,.
\lab{g-cross}
\ee
In the large $s$ limit we have the following expressions for the eigenvalues
of the matrix $\Gamma_{\rm cross}(\gamma)$:
\be
\Gamma_{+}=N\log\frac{s}{m^2} -N- \frac{2i\pi}{N} + O(\log^{-1}\frac{s}{m^2}),
\lab{Gamma+}
\ee
\be
\Gamma_{-}={\pi}^2\frac{N^2-1}{N^3}\log^{-1}\frac{s}{m^2} + O(\log^{-2}
\frac{s}{m^2})\,.
\lab{Gamma-}
\ee
The scattering amplitude can be decomposed into singlet and octet
invariant amplitudes corresponding to exchanges in the $t$-channel with
quantum numbers of the vacuum and the gluon, respectively:
\be
T_{ij}^{i'j'}=\delta_{i'j}\delta_{j'j}T^{(0)}
+t_{i'i}^{a}t_{j'j}^{a}T^{(8)}\,.
\lab{dec}
\ee
Using the one-loop expression \re{g-cross} for the matrix
$\Gamma_{\rm cross}(\gamma,g)$ we find
the following expressions for the invariant amplitudes:
\be
T^{(0)}=-\sinh\gamma(\frac{\Gamma_{-}}{\Gamma_{+}}T_{+} - T_{-})\,,
\lab{T0}
\ee
\be
T^{(8)}=2i\pi\cosh\gamma\frac{1}{\Gamma_{+}}(T_{+} - T_{-})
\lab{T8}
\ee
where
\be
T_{\pm}=\int d^2b\,\\e^{-i\vec b\vec k}\exp\left(-\Gamma_{\pm}
\int_{\lambda}^{1/b}\frac{\alpha(\tau)}{\pi}\frac{d\tau}{\tau}\right)\,.
\lab{W+-}
\ee
In the leading $\log t$ and $\log s$ approximation the result for the
invariant scattering amplitude has the standard reggeized form \ci{PT}:
\be
T^{(0)}=0  \qquad\quad  T^{(8)}=T_{Born}\left(\frac{s}{m^2}
\right)^{\alpha(t)},
\lab{rf}
\ee
where $T_{Born}=\frac{ig^2}{t}\frac{s}{m^2}$.
The nonleading $\log s$ corrections drastically change the behavior of the
scattering amplitude. Indeed, the functions $T_{\pm}$ \re{W+-}
have a Regge-like behavior,
and for $\Gamma_{+} \gg \Gamma_{-}$ we have $T_{+} \ll T_{-}$. Thus, the
high-energy behavior of the invariant scattering amplitudes
\re{T0}, \re{T8}  is dominated by
the contribution of $T_{-}$, and as a consequence the amplitude of the octet
exchange is suppressed by the factor $1/\Gamma_{+}=0(\log^{-1}(s/m^2))$
compared to the amplitude of the singlet exchange. Therefore we will start
from the main result of perturbative QCD:
\be
T^{(0)}= \sinh\gamma T_{-}
\lab{pT0}
\ee
\be
T^{(8)}=-\frac{2i\pi}{\Gamma_{+}}\cosh\gamma T_{-}.
\lab{pT8}
\ee

\section{IR renormalons in Wilson loops}
\setcounter{equation} 0
Expression \re{W+-} for the scattering amplitude  was found by summing
all large Sudakov logarithms which were artificially extracted from the
uniquely defined scattering amplitude $T_{ij}^{i'j'}$ \re{main}. As a result
the perturbative expansion for the scattering amplitude, as we will show
in this section, is not well defined, and has ambiguities associated with
IR renormalons. To restore the uniqueness of the physical quantity, the
perturbative expression should be supplemented by nonperturbative
corrections. However, we have to pay for this by introducing a new scale,
which characterizes the size of nonperturbative effects. We do not know
how to evaluate the nonperturbative corrections but we may predict in
general their structure and dependence on impact parameter $b$ by exploiting
the idea of IR renormalons.

The occurrence of infrared renormalons in the amplitude may be seen
explicitly in an ``improved'' calculation of the Wilson line (WL) expectation
value by replacing
$\alpha_{s}$ by the running coupling constant $\alpha_{s}(\vec k^2)$
\ci{kor.ster}.
As an example, we consider the one-loop calculation of  $W_1$  of Fig. 2 in
Feynman gauge, and find the structure of the nonperturbative correction for
this particular diagram.
\be
W_{1} =  4(t^a\otimes t^a)\tilde{\mu}^{4-D}
(p_{1}p_{2})\int\frac{d^Dk}{(2\pi)^D}\int_{-\infty}^{+\infty}d\alpha
\int_{-\infty}^{+\infty}d\beta\\e^ {i(p_{1}\beta+b)k}\\e^ {-i(p_{2}k)\alpha}
\frac{i\alpha_{s}(\vec k^2)}{k^2+io}\,,
\lab{W1}
\ee
where $D=4-2\varepsilon$   , $d^Dk=dk^+dk^-d^{D-2}\vec k$\,.
After integration over $\alpha, \beta$ and $k_{\pm}$ in \re{W1} we get
the vacuum
average of the WL as an integral over the gluon transverse momenta:
\be
W_{1}= 4(t^a\otimes t^a)\tilde{\mu}^{4-D}
(-i\pi \coth\gamma)\int\frac{d^{D-2}\vec k}{(2\pi)^{D-2}}
\frac{\alpha_{s}(\vec k^2)}{\vec k^2 - io}
\\e^{i\vec k\cdot \vec b}\,.
\lab{Wtr}
\ee
Notice that after integration over  $\vec k^2$, even for fixed coupling
constant this expression contains infrared poles in $\varepsilon$.
To regularize IR divergences we have introduced dimensional
regularization, with scale $\tilde{\mu}$. The IR cutoff $\lambda$ (see Sect. 2)
is neglected for this argument, and we may assume for a moment that
$\lambda \ll \Lambda$.
The cross divergence does not appear for nonzero impact parameter  $ b $,
because $b$ regularizes the gluon propagator at short distances.
Let us now review how IR renormalons appear \ci{kor.ster}.

After substitution of the relation  $\alpha_{s}(\vec k^2)=\int_{0}^{\infty}
d\sigma(\vec k^2/\Lambda^2)^{-\sigma\beta_{1}}$  into \re{Wtr}, and after
integration over transverse momenta, we get
\be
W_{1} = -(t^a\otimes t^a)\frac{1}{\pi}(-i\pi \coth\gamma)
(\pi\tilde{\mu}^{2}b^2)^{\epsilon}
\int\frac{d\sigma}{(\varepsilon + \beta_{1}\sigma)}
\frac{\Gamma(1-\varepsilon - \beta_{1}\sigma)}
{\Gamma(1+\beta_{1}\sigma)}
\left(\frac{b^2\Lambda^2}{4}\right)^{\beta_{1}\sigma}.
\lab{W2}
\ee
We notice that $\left(\frac{b^2\Lambda^2}{4}\right)^{\beta_{1}\sigma}=e^
{-\sigma/\alpha_{s}}$  with  $\alpha_{s}=\alpha_{s}(4/b^2)$ .
We thus identify the right-hand side of \re{W2} as the Borel representation
\be
\pi(\alpha_{s})=\int_{0}^{\infty} d\sigma \tilde{\pi}(\sigma)\,
\\e^{-\sigma/\alpha_{s}}
\lab{pi}
\ee
of  $\pi(\alpha_{s})\equiv W_{1}$ , with
\be
\pi(\alpha_{s})= -(t^a\otimes t^a)\frac{1}{\pi}(-i\pi \coth\gamma)
(\pi\tilde{\mu}^{2}b^2)^{\epsilon}
\int\frac{d\sigma}{(\varepsilon + \beta_{1}\sigma)}
\frac{\Gamma(1-\varepsilon - \beta_{1}\sigma)}
{\Gamma(1+\beta_{1}\sigma)}\,.
\lab{IR}
\ee
The limit $\sigma\to 0$ produces a singularity in $\varepsilon$ due to
IR divergences in \re{W1}, because we have neglected the IR cutoff $\lambda$.
Away from $\sigma=0$  we put $\varepsilon=0$ and find that the function
$\tilde{\pi}(\sigma)$  has singularities generated by $\Gamma$ -function
at  $\sigma^*= 1/\beta_{1}, 2/\beta_{1}, 3/\beta_{1}, \ldots$ . These are
the infrared renormalons. The result of integration in \re{pi} depends on
the regularization prescription.
That is, the factor $e^{-\sigma^*/\alpha_{s}}=(b^2\Lambda^2)^n$
induces an ambiguity in the scattering amplitude
at the level of the power corrections $ O(b^2\Lambda^2)^n $. This
fact, together with the observation that the infrared renormalons
come from region of small gluon momenta $(k\sim \Lambda)$ where the coupling
constant becomes large and where we expect the appearance of nonperturbative
corrections, implies that the nonperturbative corrections are
power corrections.
The first renormalon gives a contribution at the level $ O(b^2\Lambda^2) $.
Thus, for the WL to be well
defined, nonperturbative effects should contribute at the same level.
Hence the nonperturbative correction has the following form:
\be
\rho(t^a\otimes t^a)(i\pi \coth\gamma)(b^2\Lambda^2)\,,
\lab{noncor}
\ee
where $\rho$ is some parameter characterizing their size.

Let us consider the color matrix structure of nonperturbative corrections.
The direct product of the gauge group generators can be decomposed into
the sum of invariant tensors
$$
t^a_{ij} t^a_{kl} = -\frac1{2N}\delta_{ij}\delta_{kl}
+\frac12\delta_{il}\delta_{jk}.
$$
Therefore the structure  $(t^a\otimes t^a)(i\pi \coth\gamma)$, which
corresponds to the particular configuration  $W_{1}$ of Fig. 2,
reproduces the first two elements of the matrix cross anomalous
dimensions \re{g-cross}.
Notice that the structure $ i\pi \coth\gamma $ containing the
dependence on $\gamma$ is factorized after integrating over $\alpha$ and
$\beta $-Wilson line parameters and over $ k_{\pm} $. Evidently the angle and
group structure
of the one-loop diagrams of $ W_{2} $ reproduce the other two
elements of the matrix cross anomalous dimension \re{g-cross}.
Therefore we conclude that the factorized matrix structure of nonperturbative
corrections is exactly  $ \Gamma_{\rm cross}(\gamma)$ .

In summary, we may represent $W_{1}$ as a sum of perturbative and
nonperturbative terms as
\be
W_{a}(\gamma,\lambda^2b^2)=\left(\delta_{ab}+\Gamma_{\rm cross}^{ab}(\gamma)
\int_{\lambda}^{1/b}\frac{\alpha(\tau)}{\pi}\frac{d\tau}{\tau}-
\Gamma_{\rm cross}^{ab}(\gamma)
\frac{\rho}{\pi}\Lambda^2b^2\right)W_{b}(\gamma,1)\,,
\lab{1loop}
\ee
where $a=1,\, b=1,2$ and $\lambda$ is IR cutoff. We must emphasize that we
choose
$\lambda \geq \Lambda$ to be able to consider the influence of nonperturbative
effects. In general the nonperturbative parameter $\rho$ depends on
the IR cutoff $\lambda$. We recall that in this section and afterward
$\Lambda$ is fundamental QCD scale.

\section{Full scattering amplitude}
\setcounter{equation} 0
It is now natural to generalize the exponentiated expressions
\re{W+-}, \re{pT0}, \re{pT8} to include the nonperturbative corrections.
The Wilson lines \re{main1} are defined beyond perturbation theory.
In perturbation theory, however, we know the renormalization group
equation \re{RG} for the Wilson line and its solution.
That is why it is natural to consider the exponentiation of nonperturbative
corrections as well as perturbative. In fact at one-loop level, the matrix
structures of perturbative and nonperturbative corrections are the same.
The origin of this property is the following.

As was shown in the previous
section, in Feynman diagrams contributing to the Wilson loop
nonperturbative corrections come from the integration over small
transverse momenta. At the
same time, the $\gamma$ - dependence appears when one integrates over small
angles between gluon momenta and quark momenta . Since integrations over small
angles are completely independent of the integration over small transverse
momenta, the $\gamma$- dependence of both perturbative and nonperturbative
results coincides.
In the paper \ci{Kor.Kor} it was shown that higher order corrections
preserve the asymptotic behavior of the
eigenvalues of the matrix cross anomalous dimension. Together,
these arguments imply that nonperturbative corrections have $\Gamma_{-}\sim
1/\gamma$ and $\Gamma_{+}\sim  \gamma$ asymptotics. Therefore the
generalization can be performed in the same manner as the one-loop
example before \re{1loop}
\be
T_{\pm}(\vec k^2)=\int d^2b\,\\e^{-i\vec b\cdot \vec k}\exp\left(-\Gamma_{\pm}
\int_{\lambda}^{1/b}\frac{\alpha(\tau)}{\pi}\frac{d\tau}{\tau}-\Gamma_{\pm}
\frac{\rho}{\pi}\Lambda^2b^2\,
\right).
\lab{nW+-}
\ee
For small values of $b$ the scattering amplitude gets its main contribution
from the perturbative region. At large $b$ the nonperturbative corrections
become important.

For invariant amplitudes we find expressions like (2.17) and (2.18):
\be
T^{(0)}= \sinh\gamma T_{-}
\lab{nT0}
\ee
\be
T^{(8)}=-\frac{2i\pi}{\Gamma_{+}}\cosh\gamma T_{-}
\lab{nT8}
\ee
We expect that the condition $T_{+} \ll T_{-}$ is conserved. Moreover,
this condition is provided explicitly by the structure
$\frac{\rho}{\pi}\Gamma_{+}\Lambda^2b^2$ in the argument of the exponent,
because in high energy limit $\Gamma_{+} \to \infty$.

Now we come to the main problem of our paper : how to evaluate
the integral over impact parameter $\vec b$ in the expression for $T_{-}$
\re{nW+-}.
We will use the method of representation of the Fourier transformation
via a Mellin transformation, which was so elegantly applied to the Drell-Yan
process \ci{Collins.Soper}. We should mention also the papers
\ci{Jones.Wyndham.Rakow.Webber}, which contain some interesting calculations
based on Mellin transformation technique.
Let us briefly formulate this method.

If we define
\be
T_{M}(s)=\int\frac{d^2b}{b^2}(\Lambda^2b^2)^s\tilde{T}_{-}(b^2)\,,
\lab{Wm}
\ee
where $\tilde{T}_{-}(b^2)$ is the Fourier transform of $T_{-}(\vec k^2)$:
$$
\tilde{T}_{-}(b^2)=\int\frac{d^2\vec k}{{(2\pi)}^2}\,\\e^{i\vec b\cdot \vec k}
\,T_{-}(\vec k^2)\,,
$$
then the Mellin theorem gives us $T_{-}(\vec k^2)$ in the following form:
\be
T_{-}(\vec k^2)=\frac{1}{2\pi i}\frac{4\pi}{k^2}\int_{\delta-i\infty}^
{\delta+i\infty}
ds\left(\frac{\vec k^2}{4\Lambda^2}\right)^s \frac{\Gamma(1-s)}{\Gamma(s)}
T_{M}(s) \, , \qquad\quad 0 < \delta < 1  \,,
\lab{Wk-main}
\ee
where
\be
T_{M}(s)=
\int\frac{db^2}{b^2}(\Lambda^2b^2)^s
\exp\left(-\Gamma_{-}
\int_{\lambda}^{1/b}\frac{\alpha(\tau)}{\pi}\frac{d\tau}{\tau}-\Gamma_{-}
\frac{\rho}{\pi}\Lambda^2b^2
\right).
\lab{Ts-main}
\ee
Therefore we start our detailed calculation of the scattering amplitude
with \re{Wk-main} and \re{Ts-main}. We consider two cases: frozen and running
coupling constants.

\section{Scattering amplitude for a frozen coupling constant}
\setcounter{equation} 0
First we consider the situation when the coupling constant does not run.
For this case, one will be able to get an exact answer for the scattering
amplitude. Let us rewrite expression \re{Ts-main} for $T_{M}(s)$
taking into account
that $\alpha_{s}(\tau) \equiv \alpha_{s}$. We get

\be
T_{M}(s)=
\int\frac{db^2}{b^2}(\Lambda^2b^2)^s
\exp\left(\frac{\alpha_{s}\Gamma_{-}}{2\pi}
\log\lambda^2b^2 - \frac{\rho}{\pi}\Gamma_{-}b^2\Lambda^2\right).
\lab{Ts-fr}
\ee
It is convenient to denote
$$
\alpha=\frac{\alpha_{s}\Gamma_{-}}{2\pi}\,,
$$
\be
R^2=\frac{\rho}{\pi}\Gamma_{-}\Lambda^2\,.
\lab{R}
\ee
{}From the beginning let us integrate in $b$-space, and after that
concentrate our efforts on the integration over $s$.
Evidently, $T_{M}(s)$ is proportional the Gamma-function.
Substituting the expression obtained for $T_{M}(s)$ in the expression
\re{Wk-main} for the scattering amplitude, in which it is useful to shift
the integration variable $s\rightarrow s-1$ , one finds
\be
T_{-}(\vec k^2)=\frac{1}{2i\pi}\frac{\pi}{R^2}\left(\frac{\lambda^2}
{R^2}\right)^\alpha\int_{\delta-i\infty}^{\delta+i\infty}
ds\left(\frac{\vec k^2}{4R^2}\right)^s \frac{\Gamma(-s)
\Gamma(1+s+\alpha)}{\Gamma(s+1)}\,.
\lab{Wk-fr1}
\ee
Note, if we deal with only nonperturbative corrections ($\alpha=
\alpha_{s}\Gamma_{-}/2\pi \equiv 0$) the expression \re{Wk-fr1} is simplified.
Imagining the contour to envelope the right half plane
and summing over all residues of $\Gamma(-s)$ we find
\be
T_{-}(\vec k^2)=\frac{\pi}{R^2}\sum_{s=0}^{\infty} \frac{(-1)^s}
{s!}{\left(\frac{\vec k^2}{4R^2}\right)}^s=\frac{\pi}{R^2}\,\\e^
{-\frac{\vec k^2}{4R^2}}\,.
\ee
This answer coincides, of course, with the result of direct integration
over impact parameter $\vec b$ in the expression \re{nW+-} for $T_{-}$,
and sheds some light on the occurance of the Gaussian distribution on
transferred momentum.

Let us continue and recall the definition of the confluent hypergeometrical
function (CHF) by means of Mellin integral representation (see Appendix).
Taking into account the remarkable property of the CHF \re{RP},
and identifying $a=\alpha+1$,  $c=1$,  $x=\frac{\vec k^2}{4R^2}$, one obtains
\be
T_{-}(\vec k^2)=\frac{\pi}{R^2}\left(\frac{\lambda^2}{R^2}\right)^\alpha
\\e^{-x}F(-\alpha,1,x)\Gamma(1+\alpha)\,.
\lab{Wk-fr2}
\ee
A more detailed form of this expression, using definition \re{R}, is
\be
T_{-}(\vec k^2)=\frac{\pi^2}{\rho\Lambda^2\Gamma_{-}}\left(\frac{\lambda^2}
{\Lambda^2}\frac{\pi}{\rho\Gamma_{-}}
\right)^{\frac{\alpha_{s}\Gamma_{-}}{2\pi}}\\e^
{{-\frac{\vec k^2}{4\Lambda^2}\frac{\pi}{\rho\Gamma_{-}}}}
F\left(-\frac{\alpha_{s}\Gamma_{-}}{2\pi} ,\,1 ,
-\frac{\vec k^2}{4\Lambda^2}\frac{\pi}{\rho\Gamma_{-}}
\right)
\Gamma(1+\frac{\alpha_{s}\Gamma_{-}}{2\pi})\,.
\lab{Wk-fr3}
\ee
This is the exact expression for the scattering amplitude in the case
when the coupling constant does not run. Let us find the asymptotics of
\re{Wk-fr2} using the properties of the CHF \re{x<1}, \re{x>1}.
There are two limiting cases to consider: $x \ll 1$ and $x \gg 1$, and
we certainly
have a critical value of transferred momentum $k_{\rm crit}$ such that
\be
\frac{\vec k^2_{\rm crit}}{4\Lambda^2}=\frac{\rho}{\pi}\Gamma_{-}=
\rho\pi\frac{N^2-1}{N^3}\frac{1}{\log s/m^2}\,\,,
\lab{crit}
\ee
which corresponds to $x \equiv \frac{\vec k^2}{4R^2}=1$. At this value
there is a crossover between rapid Gaussian decrease over $\vec k$ in the
scattering amplitude and slower Regge dependence in $t$.
\begin{itemize}
\item{1.}
$ k \ll k_{\rm crit}$, which corresponds to $x \ll 1$.
\end{itemize}
Substituting \re{x<1} in \re{Wk-fr2} we get
\be
T_{-}(\vec k^2)=\frac{\pi}{R^2}\left(\frac{\lambda^2}{R^2}\right)^\alpha
\\e^{-x} (1-\alpha(x-\psi(1)) + \ldots),
\lab{Wk-fr4}
\ee
where $\psi(x)=d\log\Gamma(x)/dx$.
The result is a Gaussian distribution over the transferred momentum.
\begin{itemize}
\item{2.}
$ k \gg k_{\rm crit}$, which corresponds to $x \gg 1$.
\end{itemize}
Substituting \re{x>1} in \re{Wk-fr2} we get
\be
T_{-}(\vec k^2)=2\alpha_{s}\frac{\Gamma_{-}}{-\vec k^2}
\left(\frac{s}{m^2}\right)^{\frac{\Gamma_{-}}{\Gamma_{+}}\alpha(t)}
\frac{\Gamma(1+\alpha)}{\Gamma(1-\alpha)}\,.
\lab{Wk-fr5}
\ee
The result has the standard reggeized form with Regge trajectory
$ \alpha(t)=-\frac{\alpha_{s}}{2\pi}N\log\frac{-t}{4\lambda^2}$
and $ -t=\vec k^2$.


\section{The scattering amplitude as an asymptotic series:
Running coupling constant.}
\setcounter{equation} 0
Let us substitute the expression for the running coupling constant,
$\alpha_{s}(\tau)=1/(\beta_{1}\log\frac{\tau^2}{\Lambda^2})$ in
\re{Ts-main} and integrate over $\tau$.
Then
\be
T_{M}(s)=\int\frac{db^2}{b^2}(\Lambda^2b^2)^s
\exp(-\frac{\Gamma_{-}}{2\pi\beta_{1}}\log\frac{\log\frac{1}
{\Lambda^2b^2}}{\log\frac{\lambda^2}{\Lambda^2}}-\frac{\rho}{\pi}
\Gamma_{-}b^2
\Lambda^2)\,.
\lab{Ts-ex}
\ee
To evaluate a double integral like \re{Wk-main}, Collins and Soper proposed
in \ci{Collins.Soper}
a saddle point approximation. For the Drell-Yan process this method is
applicable because in the high energy limit the nonperturbative corrections
play no role, and the saddle point lies in the perturbative region. In
quark-quark scattering, however, the application of this method
is doubtful from the beginning because the factor $\Gamma_{-}$, which
appears in
the argument of the exponent in \re{Ts-ex}, approaches zero in the high
energy limit.
Moreover, the position
of the saddle point is very sensitive to the nonperturbative parameter $\rho$.
The saddle point may lie in perturbative region or out of it.
That is why we suggest a  method of evaluating the integral
based on the sum of the contributions from all  poles in $s$.

It is useful now to introduce new variables,
$$
n=\frac{\Gamma_{-}}{2\pi\beta_{1}}\,,
$$
\be
R^2=\frac{\rho}{\pi}\Gamma_{-}\Lambda^2\,,
\lab{def2}
\ee
$$
\theta=\log\frac{\lambda^2}{\Lambda^2}\,.
$$
Using an $\alpha$-representation for $\frac{1}{\left(\log\frac{1}
{\Lambda^2b^2}\right)^n}$
and integrating over $b$ in \re{Ts-ex}, one finds
\be
T_{M}(s)=\frac{\theta^n}{\Gamma(n)}\int_{0}^{i\infty}\left(\frac{\Lambda^2}
{R^2}\right)^{s+\alpha}\alpha^{n-1}\Gamma(s+\alpha)d\alpha\,.
\lab{I1-ex}
\ee
The integral $T_{M}(s)$ is not well defined at $i\infty$ because the
expression $\frac{1}{\left(\log\frac{1}{\Lambda^2b^2}\right)^n}$ is
divergent at $b^2=1/\Lambda^2$. This is a consequence of the divergence
in the running coupling constant at $\vec k^2=\Lambda^2$.
Taking this remark into account, let us go ahead and try to represent the
integral as an asymptotic series. We also find the condition under which the
asymptotic series can be approximated by the first terms and show that it
is reasonable.

In the expression \re{Wk-main} for $T_{-}(\vec k^2)$ we shift variable $s$
to $s - 1$ and substitute the result \re{I1-ex} for $T_{M}(s)$.
We get,
\be
T_{-}(\vec k^2)=\frac{1}{2i\pi}\frac{\pi}{R^2}\frac{\theta^n}{\Gamma(n)}
\int_{-i\infty}^{i\infty}ds\left(\frac{\vec k^2}{4R^2}\right)^s\frac
{\Gamma(-s)}{\Gamma(s+1)}\int_{0}^{i\infty}\Gamma(1+s+\alpha)\left(\frac
{\Lambda^2}{R^2}\right)^\alpha\alpha^{n-1}d\alpha\,.
\lab{Wk1-ex}
\ee
We notice that this expression is similar to \re{Wk-fr1} except for the
integration over $\alpha$.
Using the definition of the CHF
by means of Mellin integral representation \re{CHF},
taking into account the property \re{RP} of the CHF
and identifying $a=\alpha+1$,  $c=1$,  $x=\frac{\vec k^2}{4R^2}\,$,
one  obtains
\be
T_{-}(\vec k^2)=\frac{\pi}{R^2}\frac{\theta^n}{\Gamma(n)}\\e^{-\frac{\vec k^2}
{4R^2}}\int_{0}^{i\infty}\Gamma(1+\alpha)F(-\alpha,1,x)
\left(\frac{\Lambda^2}{R^2}\right)^\alpha\alpha^{n-1}d\alpha\,.
\lab{Wk2-ex}
\ee
As in Section 5 we concentrate our attention on the two limiting cases.
The critical value of the transferred momentum is determined as before
by \re{crit}.
\begin{itemize}
\item {1.}
$ k \ll k_{\rm crit}$, which corresponds to $x \ll 1$.
\end{itemize}
Substituting \re{x<1} in \re{Wk2-ex} we get
$$
T_{-}(\vec k^2)=\frac{\pi}{R^2}\frac{\theta^n}{\Gamma(n)}\\e^{-\frac{\vec k^2}
{4R^2}}\int_{0}^{i\infty}\Gamma(1+\alpha)(1-x\alpha+\ldots)
\left(\frac{\Lambda^2}{R^2}\right)^\alpha\alpha^{n-1}d\alpha
$$
\be
=\frac{\pi}{R^2}\frac{\theta^n}{\Gamma(n)}\\e^{-\frac{\vec k^2}{4R^2}}
 \left(J_{1} -xJ_{2} +\ldots\right)\,,
\ee
where
\be
J_{1}=\int_{0}^{i\infty}\Gamma(1+\alpha)
      \left(\frac{\Lambda^2}{R^2}\right)^\alpha\alpha^{n-1}d\alpha\,,
\lab{J1}
\ee
\be
J_{2}= \int_{0}^{i\infty}\Gamma(1+\alpha)
       \left(\frac{\Lambda^2}{R^2}\right)^\alpha\alpha^{n}d\alpha\,.
\ee
One can easily determine corrections by keeping more terms in the
expansion of the CHF.
Consider the integral $J_{1}$ \re{J1} and represent it as an asymptotic series.
In this case, one can turn the contour from $(0, i\infty)$ to
$(0, \infty)$ or change variable $\alpha$ to $i\alpha$.
After expanding the Gamma function $\Gamma(1+\alpha)$ as a series
in powers of  $\alpha$ and integrating  over $\alpha$ we get
\be
J_{1}=\sum_{r=0}^{\infty}\frac{\Gamma^{(r)}(1)}{r!}\frac{\Gamma(n+r)}
{\left(\log{\frac{R^2}{\Lambda^2}}\right)^{n+r}}
=\frac{\Gamma(n)}{\left(\log{\frac{R^2}{\Lambda^2}}\right)^{n}}
\left(1+ \frac{n\psi(1)}{\log{\frac{R^2}{\Lambda^2}}} +
\frac{n(n+1)(\psi^\prime(1)+\psi^2(1))}{\left(\log{\frac{R^2}{\Lambda^2}}
\right)^{2}} + \ldots\right)\,.
\lab{J11}
\ee
We now have the expression \re{J11} for $J_{1}$ in terms of an asymptotic
series. The first terms of this series yield accurate value for
$J_{1}$ when
\be
|\log\frac{R^2}{\Lambda^2}| = |\log \frac{\vec k_{\rm crit}^2}{4\Lambda^2}|
\gg 1
\lab{con1}
\ee
Note that in high energy limit  $\,n \rightarrow 0$ and therefore we
didn't include  $\,n$ in the condition \re{con1}.
$J_{2}$ differs from $J_{1}$ only by an additional power of $\alpha$ in the
integral.
It means that the contributions from $J_{2}$ are suppressed compared to
those from $J_{1}$. Evidently the same argument can be applied to $J_{i}=
\int_{0}^{i\infty}\Gamma(1+\alpha)
\left(\frac{\Lambda^2}{R^2}\right)^\alpha\alpha^{n-2+i}d\alpha, $
$i=3,4$,... Therefore
we conclude that the scattering amplitude can be approximated under the
condition \re{con1} by the first terms of the asymptotic series:
\be
T_{-}(\vec k^2)=\frac{\pi}{R^2}\frac{\theta^n}{\left(\log{\frac{R^2}
{\Lambda^2}}\right)^{n}}\\e^{-\frac{\vec k^2}{4R^2}}
\left(1+ \frac{n}{\log{\frac{R^2}{\Lambda^2}}}(\psi(1)-x+ O(x^2))+
O\left(\frac{1}{\left(\log{\frac{R^2}{\Lambda^2}}\right)^{2}}\right) +
\ldots \right).
\ee
A more detailed form, using definition \re{def2}  is
\be
T_{-}(\vec k^2)=\frac{\pi^2}{\rho\Lambda^2\Gamma_{-}}\\e^{
{-\frac{\vec k^2}{4\Lambda^2}\frac{\pi}{\rho\Gamma_{-}}}}
\exp\left(-\frac{\Gamma_{-}}{2\pi\beta_{1}}\log{\frac{\log
\frac{\rho\Gamma_{-}}{\pi}}
{\log\frac{\lambda^2}{\Lambda^2}}}\right)
\left(1 - \frac{1}{2\rho\beta_{1}}\frac{1}{\log{\frac{\rho\Gamma_{-}}{\pi}}}
\frac{\vec k^2}{4\Lambda^2} +\ldots\right).
\lab{gauss}
\ee
Therefore we have the Gaussian distribution over transferred momentum.
Let us investigate in more detail the first exponent in \re{gauss}.
Using the expression \re{Gamma-} for the eigenvalue $\Gamma_{-}$ one
can rewrite it as
\be
\left(\frac{s}{m^2}
\right)^{-\frac{\vec k^2}{4\Lambda^2}\frac{1}{\pi\rho}\frac{N^3}{N^2-1}}
\ee
This means that by increasing energy $s$ the bulk of the diffraction peak,
which is concentrated for
$$
t < \frac{2\Lambda^2}{\pi\rho}\frac{N^2-1}{N^3}\frac{1}{\log{s/m^2}}\,,
$$
becomes narrower. This phenomena is called shrinkage, which we have thus
derived from IR renormalon analysis. The slope of the soft gluon
trajectory is
\be
\alpha'=\frac{1}{4\Lambda^2}\frac{1}{\pi\rho}\frac{N^3}{N^2-1}\,,
\lab{slope}
\ee
and we can estimate the value of nonperturbative parameter $\rho$
using the experimental result for the slope: $ \alpha'=0.25 \rm Gev^{-2}$
\ci{Land}.
It turn out that $\rho \sim 27$.

Let us now substitute \re{gauss} into expression \re{nT0} for invariant
vacuum amplitude and evaluate the
differential cross section as
$ d\sigma/dt=1/s^2|T^{(0)}|^2$. Taking into account that at high energy
$2\sinh\gamma \to s/m^2$ and factor
$1/\Gamma_{-}\sim \log(s/m^2) $, we find
that the differential cross section is given by:
\be
\frac{d\sigma}{dt} \sim \log^2\left(\frac{s}{m^2}\right)
\left(\frac{s}{m^2}\right)^{2\alpha't}
\exp\left(-\frac{\Gamma_{-}}{\pi\beta_{1}}\log{\frac{\log
\frac{\rho\Gamma_{-}}{\pi}}
{\log\frac{\lambda^2}{\Lambda^2}}}\right)
\lab{fro}
\ee
Notice that the exponential factor in the expression \re{fro} has a
very interesting feature  : the nonperturbative parameter $\rho$
has penetrated into the perturbative expansion. This factor
should determine the intercept of the soft pomeron. However, based on
our analysis we can only estimate it qualitatively. All that we can say is
that the intercept depends on the nonperturbative parameter $\rho$,
depends slightly on the energy through $\Gamma_{-}$, and at high energy
is close to 1. Moreover, the cross section \re{fro} cannot grow faster than
$\log^2 s$ at the limit $s\to \infty$. We conclude thus that
our soft pomeron satisfies the Froissart bound.

\begin{itemize}
\item{2.}
$ k \gg k_{\rm crit}$, which corresponds to $x \gg 1$.
\end{itemize}
Substituting \re{x>1} in \re{Wk2-ex} we get the following
expression for $T_{-}(\vec k^2)$\,,
\be
T_{-}(\vec k^2)=-\frac{4\pi}{\vec k^2}\frac{\theta^n}{\Gamma(n)}
\int_{0}^{i\infty}\Gamma^2(1+\alpha)\frac{\sin\pi\alpha}{\pi}
\left(\frac{4\Lambda^2}{\vec k^2}\right)^\alpha\left(1+\frac{(1+\alpha)^2}
{x}+ \ldots\right)\alpha^{n-1}d\alpha\,.
\ee
Expanding $ \Gamma^2(1+\alpha) $ and $\sin\pi\alpha$ in
powers of $\alpha$ and integrating over $\alpha$ we get
an asymptotic series
\be
T_{-}(\vec k^2)=-\frac{2\Gamma_{-}}{\beta_{1}}\frac{1}
{\vec k^2\log\frac{\vec k^2}{4\Lambda^2}}
\exp\left(-\frac{\Gamma_{-}}{2\pi\beta_{1}}\log{\frac{\log
\frac{\vec k^2}{4\Lambda^2}}
{\log\frac{\lambda^2}{\Lambda^2}}}\right)
\left(1 + \frac{\rho\Gamma_{-}}{\pi}\frac{1}{\frac{\vec k^2}{4\Lambda^2}} +
\frac{2\psi(1)}{\log\frac{\vec k^2}{4\Lambda^2}}
(1 + \frac{\Gamma_{-}}{2\pi\beta_{1}}) +
\ldots\right)\,,
\lab{regge}
\ee
which is valid for
\be
|\log\frac{\vec k^2}{4\Lambda^2}| \gg 1.
\lab{con2}
\ee
The lowest order is independent of $\rho$.
This  means that the nonperturbative corrections can be neglected
in this region.
The formula \re{regge} is a generalization of the standard reggeized form.

The critical value of momentum transfer $k_{\rm crit}$ \re{crit} depends
on the energy $s$. Taking into account the formula for the slope  \re{slope}
we find   $ \vec k_{\rm crit}^2=1/(\alpha'\log s/m^2)$.
Therefore by increasing the energy, the value of $k_{\rm crit}$ decreases. We
conjecture that this feature together with the change in the behavior
of the scattering amplitude at $k_{\rm crit}$ describes the crossover region
(probably the dip region) in the pp differential cross section.
Since the scale of nonperturbative effects is typically 1 Gev (see, for
example \ci{DGP}), we identify $k_{\rm crit}$ with 1 Gev. Then the condition
\re{con1} is valid.

To show the consistency between the expressions for the
scattering amplitude in case of frozen and running coupling constant one
can reduce \re{regge} to \re{Wk-fr5} and \re{gauss} to \re{Wk-fr4}.
by taking the limit $\beta_{1}\to 0$.
\section{Summary and conclusions}
We began by expressing the scattering amplitude in terms of path ordered
exponentials, Wilson lines, which admit a nonperturbative definition.

We then showed that the resummation of soft gluon perturbative
corrections, leads to ambiguities in the perturbative
series associated with IR renormalons.

By examining the IR renormalon structure we predicted the form of
nonperturbative corrections, using
the idea that the ambiguity of the perturbative series is compensated
by an ambiguity in the determination of nonperturbative correction.
We thus observed that nonperturbative corrections are power corrections in
the impact variable $b$.
We restored the uniqueness of the scattering amplitude adding by ``hand''
the dominant nonperturbative corrections, which are parameterized by a new
scale.

We went on to investigate the interplay between both kinds of corrections, and
found the critical value of the transferred momentum \re{crit} at which
the behavior of the scattering amplitude is drastically changed.
In each region we represented the scattering amplitude as an
asymptotic series. For $k \ll k_{\rm crit}$ nonperturbative effects play the
main role and this region of momentum transfer corresponds to the soft
pomeron (See \re{gauss} and \re{fro}), which preserves the Froissart
bound. For $ k \gg k_{\rm crit}$ the large perturbative
Sudakov corrections are dominant (See \re{regge}).

It is naturally to expect that the nonperturbative parameter $\rho$
can be expressed as an expectation value of a nonlocal operator.
Indeed, the expression
$ \frac{\partial}{\partial b^2}W_{ij}^{i'j'}(\gamma, \vec b^2\lambda^2)$
must be related in some manner with parameter $\rho$. In \ci{kor.ster}
this connection was found explicitly for the Drell Yan process.
However, in case of q-q scattering
we have additional color structures and mixing of the Wilson lines, and,
we leave the discovery of a similar relation to future investigation.

We may generalize our considerations to study high energy elastic
quark-antiquark scattering. An antiquark can be treated as quark moving
backwards in time with opposite velocity. All the above results for
quark-quark scattering amplitude can be applied, provided that we replace
the angle $\gamma$ between quark velocities by $\gamma \to i\pi - \gamma$
in $\Gamma_{\rm cross}(\gamma)$. Then in high energy limit $(\gamma \gg 1)$
the asymptotics of the eingenvalues of the matrix $\Gamma_{\rm cross}$
do not change. Therefore the soft pomeron contributes equally to the
pp and $\bar{\rm p}$p total cross sections.

In our approach we took into account the contributions from all diagramms
describing the interaction of quarks with very soft gluons.
The basic of BKFL pomeron is harder gluons $( t \sim m^2)$.
We hope that our paper will help to understand how the two pomerons,
soft and hard, are related to each other.

Finally, to identify the physical consequences of the resummed nonperturbative
corrections, we must embed
q-q scattering in p-p scattering and conduct an additional investigation.

\bigskip\par\noindent{\Large {\bf Acknowledgments}}\par\bigskip\par

\noindent
It is a pleasure to thank G. Sterman for his patience, his support, and
for innumerable and instructive conversation. I am grateful to
G. Marchesini for interesting and helpful discussions.

\bigskip

\appendix{A}

Here we review the main properties of the confluent hypergeometrical function
(CHF) \ci{chf} that are used above.
The definition of the CHF by means of a Mellin integral representation is
\be
F(a,c,-x)=\frac{1}{2\pi i}\frac{\Gamma(c)}{\Gamma(a)}
\int_{\delta-i\infty}^{\delta+i\infty}
\frac{\Gamma(-s)\Gamma(a+s)}{\Gamma(s+c)}x^sds\,.
\lab{CHF}
\ee
We use the following properties:

1. The continuation formula of Ernst Kummer
\be
F(a,c,-x)=\\e^{-x}F(c-a,c,x)\,,
\lab{RP}
\ee

2. The asymptotics of CHF

For $ x \ll 1$
\be
F(-\alpha,1,x)=1-\alpha x+\frac{\alpha(\alpha-1)}
{2}\frac{x^2}{2}+\ldots\,,
\lab{x<1}
\ee

For $ x \gg 1 $
\be
F(-\alpha,1,x)=\\e^{x}\frac{x^{-\alpha-1}}{\Gamma(-\alpha)}(1+\frac
{(1+\alpha)^2}{x}+ \ldots)\,.
\lab{x>1}
\ee
\bb{99}
\bi{Regge}
      P.D.B. Collins, {\it ``An introduction to Regge theory and high energy

      physics''\/}, Cambridge Univ. Press (1977) 445.
\bi{BKFL}
      E.A. Kuraev, L.N. Lipatov and V.S. Fadin, Sov.Phys.JETP
      44 (1976) 443-451; 45 (1977) 199;
      Ya.Ya. Balitskii and L.N. Lipatov, Sov.J.Nucl.Phys. 28 (1978) 822.
\bi{Cheng.Wu}
      H. Cheng and T.T. Wu, {\it ``Expanding Protons: Scattering at
      High Energy''\/}, (MIT Press, Cambridge, Massachusetts, 1987).
\bi{Lip}
       L.N. Lipatov, JETP Lett. 59 (1994) 596.
\bi{FK}
       L.D. Faddev and G.P.Korchemsky, Phys. Lett. B342 (1995) 311.
\bi{K.BA for Pomeron}
       G.P.Korchemsky, Nucl. Phys. B443 (1995) 255.
\bi{Landshoff. Two Pomerons}
       P.V. Landshoff, hep-ph/9410250; hep-ph/9311201; hep-ph/9505396.
\bi{Bertini.Giffon.Predazzi}
       M. Bertini, M. Giffon, E. Predazzi, hep-ph/9501254
\bi{Donnachie.Landshoff.old}
       A. Donnachie and P.V. Landshoff, Nucl. Phys. B231 (1984) 189;
       Nucl. Phys. B244 (1984) 322.
\bi{nonpert.cor.}
       P.V. Landshoff and O. Nachtmann, Z.Phys. C35 (1987) 405;
       J.R. Cudell and B.U. Nguyen, Nucl. Phys. B420 (1994) 669.
\bi{Nachtmann}
       O. Nachtmann, preprint HD-THEP-94-42, hep-ph/9411345.
\bi{Verlinde}
       H. Verlinde and E. Verlinde, preprint PUPT-1319, (hep-th/9302104)
\bi{t'Hooft}
       G. 't Hooft, in {\it ``The Ways Of Subnuclear Physics''\/},
       Erice 1977, Ed.A.
       Zichichi(plenum, New York, 1977), p. 943; B. Lautrup, Phys. Lett.
       69B(1977) 109; G.Parisi, Phys.Lett. 76B(1978) 65; Nucl. Phys. B150
       (1979) 163; F. David, Nucl.Phys. B234 (1984) 237; Nucl.Phys. B263
       (1986) 637.
\bi{Mueller}
       A.H. Mueller, Nucl. Phys. B250 (1985) 327;
       in {\it ``QCD 20 years later''\/}, Aachen, 1992, Eds. P.M. Zerwas
       and H.A. Kastup
       (World scientific, Singapore, 1993).
\bi{kor.ster}
       G.P. Korchemsky and G. Sterman,  Nucl. Phys. B437 (1995) 415.
\bi{CS}
       H. Contopanagos and G. Sterman, Nucl. Phys. B419 (1994) 77.
\bi{Bigi}
       I.I. Bigi, M.A. Shifman, N.G. Uraltsev and A.I. Vainshtein,
                                   Phys. Rev. D50 (1994) 2234.
\bi{Neubert}
       M. Neubert and C.T. Sachrajda, preprint CERN-TH.7312/94
                                   (hep-ph/9407394).
\bi{Manohar.Wise}
       A.V. Manohar and M.B. Wise, Phys. Lett. 344B (1995) 407.

\bi{Dokshitzer.Webber}
       Yu.L. Dokshitzer and B.R. Webber, preprint Cavendish-HEP-95/2
                                   (hep-ph/9504219)
\bi{Kor.Rad}
       G.P. Korchemsky and A.V. Radyshkin, Phys. Lett. B171 (1986) 459;
       Sov.J.Nucl. Phys. 45 (1987) 127, 910.
\bi{Polyakov}
       A.M. Polyakov, {\it ``Gauge fields and strings''\/}, (Harwood Academic
       Publishers, New York, 1987).
\bi{Spinfactor}
       I.A. Korchemskaya and G.P. Korchemsky, Phys. Lett. B257 (1991) 125;
       J. Phys. A24 (1991) 4511.
\bi{Korchemsky}
       G.P.Korchemsky, Phys. Lett. B325 (1994) 459.
\bi{Nach.old}
       O. Nachtmann, Ann. Phys. 209 (1991) 436.
\bi{Lip.quaseilastic}
       L.N. Lipatov, Nucl. Phys. B309 (1988) 379;
       in {\it ``Perturbative QCD''\/}, ed. A.H. Mueller
       (World scientific, Singapore, 1989).
\bi{Sotirop.Sterman}
       M.G.Sotiropoulos and G. Sterman, Nucl. Phys. B419 (1994) 59.
\bi{Kor.Kor}
       I.A. Korchemskaya and G.P. Korchemsky, Nucl. Phys. B437 (1995) 127.
\bi{Mul.onium}
       A.N. Mueller and Bimal Patel, Nucl. Phys. B425 (1994) 471.
\bi{Brandt}
       R.A.Brandt, F. Neri and M.-A. Sato, Phys. Rev. D24 (1981) 879;
       R.A.Brandt, A. Gocksch, M.-A. Sato, and F. Neri,
       Phys. Rev. D26 (1982) 3611.
\bi{PT}
       H.T. Nieh and Y.-P. Yao, Phys. Rev. Lett. 32 (1974) 1074;
                                Phys. Rev. D13 (1976) 1082;
       B.M. McCoy and T.T. Wu,  Phys. Rev. Lett. 35 (1975) 604;
                                Phys. Rev. D12 (1975) 3257;
       L. Tyburski, Phys. Rev. D13 (1976) 1107;
       C.Y. Lo and H. Cheng, Phys. Rev. D13 (1976) 1131.

\bi{Collins.Soper}
       J.C. Collins and D.E. Soper, Nucl. Phys. B197 (1982) 446.
\bi{Jones.Wyndham.Rakow.Webber}
       H.F. Jones and J. Wyndham, J. Phys. A14 (1981) 1457;
       P.E.L. Rakow and B.R. Webber, Nucl. Phys. B187 (1981) 254.
\bi{DGP}
       P. Desgolard, M. Giffon, E. Predazzi, Z. Phys. C63 (1994) 241.
\bi{Land}
       P.V. Landshoff, hep-ph/9505254
\bi{chf}
       R.B. Dingle, {\it ``Asymptotic expansion: their derivation and
       interpretation''\/}, (Academic Press, London and New York, 1973)

\eb

\newpage
\figures

\begin{center}
\unitlength=0.50mm
\linethickness{0.4pt}
\begin{picture}(169.00,43.00)
\put(10.00,10.00){\vector(1,1){30.00}}
\put(135.00,10.00){\line(1,1){14.92}}
\put(150.00,25.00){\vector(-1,1){15.02}}
\put(166.00,10.00){\line(-1,1){15.09}}
\put(151.00,25.00){\vector(1,1){15.05}}
\put(7.00,43.00){\makebox(0,0)[cc]{$i'$}}
\put(43.00,43.00){\makebox(0,0)[cc]{$j'$}}
\put(132.00,43.00){\makebox(0,0)[cc]{$i'$}}
\put(7.00,7.00){\makebox(0,0)[cc]{$j$}}
\put(43.00,7.00){\makebox(0,0)[cc]{$i$}}
\put(132.00,7.00){\makebox(0,0)[cc]{$j$}}
\put(169.00,7.00){\makebox(0,0)[cc]{$i$}}
\put(169.00,43.00){\makebox(0,0)[cc]{$j'$}}
\put(25.00,2.00){\makebox(0,0)[cc]{$(a)$}}
\put(151.00,2.00){\makebox(0,0)[cc]{$(b)$}}
\put(24.00,26.00){\vector(-1,1){13.96}}
\put(40.00,10.00){\line(-1,1){13.96}}
\put(-15,25){\makebox(0,0)[cc]{$(W_1)^{i'j'}_{ij}=$}}
\put(110,25){\makebox(0,0)[cc]{$(W_2)^{i'j'}_{ij}=$}}
\end{picture}

\bigskip

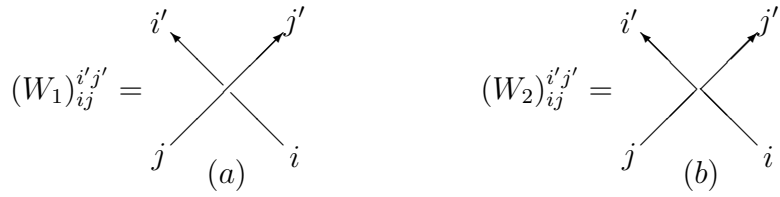
Fig.1: Integration paths (a) and (b) entering into the definition of
the Wilson lines $W_1$ and $W_2$, respectively.
\end{center}

\vspace*{10mm}

\begin{center}
\unitlength=0.50mm
\linethickness{0.4pt}
\begin{picture}(45.00,41.00)
\put(10.00,10.00){\vector(1,1){30.00}}
\put(24.00,26.00){\vector(-1,1){13.96}}
\put(40.00,10.00){\line(-1,1){13.96}}
\bezier{8}(17.00,17.00)(10.00,25.00)(17.00,33.00)
\end{picture}

\bigskip

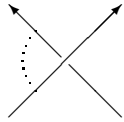
Fig.2: One-loop diagram contributing to the line function $W_1$.
Solid line represents the integration path, dotted lines denote
gluons.
\end{center}


\end{document}